\documentstyle[sprocl,epsfig]{article}

\begin{document}
\title{HARD THERMAL LOOPS FROM TRANSPORT PROCESSES}
\author{BERNDT M\"ULLER \\
	Department of Physics, Duke University \\
	Durham, NC 27708-0305, USA\\}

\maketitle 
\bigskip

\abstracts{This lecture is dedicated to the memory of Peter Carruthers, whose remarks at the previous AUP Workshop in June 1996 triggered the idea for this work.}

\section{Introduction}

Because the running coupling $g(T)$ of QCD decreases as a function of 
temperature $T$, the SU(3) color gauge theory becomes weakly coupled in the 
high temperature limit.  Numerical simulations of the lattice gauge theory with dynamical quarks have shown that the equation of state $\epsilon (T)$ makes a transition from the low-temperature phase, where the effective degrees of freedom are hadrons, to a high temperature phase, where quarks and gluons can be excited independently.  The transition temperature is $T_c \approx 150$ MeV.

Already in perturbation theory one finds that the quasiparticle excitations in the high-temperature phase are not massless.  At low momenta, $k < gT \ll T$ (we are here considering the weak coupling limit), there are three propagating collective modes with frequency 
$$
\omega(k)^2 \,\stackrel{k\to 0}{\longrightarrow}\, \omega^2_p 
= \frac{1}{3} (1+ N_f /2) g^2T^2,
$$
the longitudinally and transversely polarized plasmons.  At high momentum, the longitudinal mode is exponentially suppressed by its form factor; the transversely polarized modes carry an effective mass $m^{*2}_g 
= \frac{3}{2} w^2_p$. 

This dynamically generated mass (self energy) is essential for the avoidance of infrared divergences which first occur at order $g^4$ in the equation of state.  Incorporating the leading high-temperature contribution to the one gluon-loop diagram into effective gauge propagators and vertices, one obtains a consistent, i.e. gauge invariant,  perturbative expansion of the partition function up to order $g^5 \ln g$.  Due to remaining infrared divergences attributable to static magnetic gauge interactions, the next term of order $g^6$ cannot be calculated perturbatively.

Within the framework of perturbation theory, this divergence is seen to arise from the lack of a screening in the sector of time-independent transverse gauge fields at finite temperature.  In a more formal context, the divergence is a result of the dimensional reduction of the high-temperature gauge theory, which effectively becomes a three-dimensional theory as far as static quantities are concerned.  The three-dimensional gauge theory  is strongly coupled and confining, as revealed by the continued presence of an area law for space-like Wilson loops in the four-dimensional gauge theory at high temperature.
In other words, the physics at length scales of order $(g^2T)^{-1}$ and larger remains nonperturbative and requires numerical calculations. 

\section{Classical Color Plasma Description of Hard Thermal Loops}

The effective action describing the long-distance dynamics of the thermal gauge theory is well known:\cite{BP90,TW90,BI93}
\begin{equation} 
   \Gamma 
=  \Gamma_0 +\Gamma_{\rm HTL} 
= -\frac{1}{2} {\rm tr} F_{\mu\nu} F^{\mu \nu }
 - 3\omega^2_p \int \frac{ d^2v}{4\pi}{\rm tr} F_{\mu \lambda} 
\frac{ v^\lambda v^\nu}{(v\cdot D)^2}  F_\nu^\mu.
\end{equation}
where $v^\mu = (1, \hat{v})$ is a light-like unit vector and $D_\mu$ 
denotes the 
gauge-covariant derivative.  The second part of the action, $\Gamma_{\rm HTL}$, describes the contribution due to ``hard thermal loops'', i.e. one-loop diagrams where the gauge bosons running around the loop carry at least thermal momenta $(k \geq T)$.  The action (1) generates effective $n$-point interactions with $ n \geq 2$, in addition to the fundamental terms with $n = 2, 3, 4$. 

In spite of its formal simplicity, $\Gamma_{\rm HTL}$ poses a number of non-trivial calculational problems: 
\begin{quote} 
$\bullet$  The effective action is highly nonlocal in space and time. 

$\bullet$  In Minkowski space the effective action acquires an imaginary part describing the damping of space-like fields. 

$\bullet$ To avoid double counting, soft momenta should be excluded from the hard thermal loops, but this should be done in a gauge invariant way. 
\end{quote}
 All these problems can be solved simultaneously by replacing $\Gamma_{\rm HTL}$ with a thermal bath of classical, non-abelian charged particles that interact with the soft gauge fields.   This is possible because the physics of hard thermal loops is effectively classical: the thermal parts of loop diagrams are really tree diagrams, because the propagator in at least one section of the loop is on-shell. 

The equations describing the motion of classical, colored point particles in 
an external gauge field were first given by Wong:\cite{Wong70}
\begin{eqnarray} 
    m\dot{x}^\mu 
&=& p^\mu \\
  m \dot{p}^\mu 
&=& gQ^aF^{a\mu \nu} p_\nu \\
  m\dot{Q}^a 
&=& -gf^{abc} p^\mu A^b_\mu Q^c . 
\end{eqnarray}
The second equation describes the effect of the non-Abelian Lorentz force, while the last equation describes the precession of the color charge vector as the particle moves through the field.  This equation is really nothing else than an expression of the different definition of the color coordinate system 
at different points in space.  The equations do not describe effects of spin; 
these interactions are irrelevant in the HTL limit.  If, instead of considering individual point particles, one considers the evolution of a continuous distribution of such particles in phase space, one obtains the non-Abelian transport equation first derived by Heinz:\cite{Heinz83} 
\begin{equation} 
p^\mu\left[ 
  \frac{\partial}{\partial x_ \mu} 
  - gQ^a F^a_{\mu \nu} 
  \frac{\partial}{\partial p_\nu} 
  - gf^{abc} A^b_\mu Q^c 
  \frac{\partial}{\partial Q^a} \right] 
  f(x, p, Q) = 0, 
\end{equation}
which must be considered together with the Yang-Mills equation 
\begin{equation}  
  D_\mu F^{\mu \nu} 
= \frac{g}{m} \int dpdQ \; p^\nu f(x, p, Q) 
\equiv j^\nu (x). 
\end{equation}
The particles are on-shell, i.e. $f(x, p, Q) \sim \delta (p^2-m^2) \delta 
(Q^aQ^a - Q^2).$  The transport equation (5) is 
gauge covariant, i.e. it remains unchanged under a gauge transformation 
\begin{eqnarray} 
  A_\mu 
&\to & UA_\mu U^{-1} - \frac{1}{g} U\partial_\mu U^{-1}, \\
   f(x, p, Q) 
&\to & f(x, p, UQU^{-1}). 
\end{eqnarray} 
By considering the response of (5, 6) to a small external field $A_\mu$, one can show\cite{KLLM94} that perturbations around a thermal distribution 
\begin{equation} 
f_{\rm eq} (x, p, Q) 
= (e^{\beta_{ \mu} p^\mu} - 1)^{-1}
\end{equation}
lead to a set of equations that exactly mirror the HTL effective action (1). 
This can easily be understood in terms of the graphical representation of the one--loop contribution to the action of an ensemble of particles propagating in an external gauge field: 
$$
\centerline{\mbox{\epsfig{file=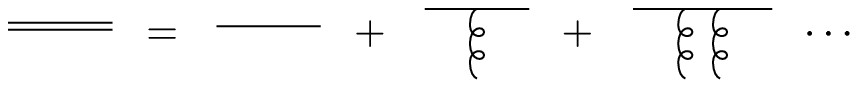,width=.80\linewidth}}}
$$
Using $p^\mu = m p^\mu$ and 
$ f(x, p, Q) = f_{\rm eq} (p) + \delta f(x, p, Q)$, one finds 
\begin{equation} 
  D^{ab}_\mu F^{b\mu \nu}
= g\int dpd Q \; Q^a v^\nu \delta f(x, p, Q) 
\equiv g\int dp \; v^\nu \delta f^a (x, p) 
\end{equation} 
together with the explicit solution of (5) in the form 
\begin{equation} 
 \delta f^a(x, p) 
= -g \frac{\partial f_{\rm eq}} {\partial |p|} 
  \int ^\infty_0 du U^{ab} \left( 
    {\vec x}, 
    {\vec x} 
- u {\vec v} \right) \cdot 
    {\vec E}^b 
   ({\vec x} 
- u {\vec v}), 
\end{equation}
where $U^{ab} ({\vec x}, {\vec y})$ denotes the path-ordered link operator between ${\vec x}$ and 
${\vec y}$.  The non-local nature of this solution is apparent. Retaining the differential form of (5) and solving this equation by means of test particles 
\begin{equation} 
f(x, p, Q) 
= \sum_i \int \delta (x^\mu - x^\mu _i (\tau)) \delta 
  (p^\mu - p^\mu _i (\tau)) \delta (Q^a - Q^a_i (\tau))d\tau
\end{equation}
is more practical.  In the limit of an infinite number of such test charges, obeying Wang's equations (2-4), $(12)$ provides an exact solution of (5). 

One additional modification must be made in practice, in order to obtain a gauge covariant system of equations: the Yang-Mills equation (6) has to be formulated on a spatial lattice.\cite{HM96}  For the gauge fields, this technique is well known since the work of Wilson and Kogut and Susskind, the lattice definition of the color source term involves some subtleties.  Essentially, the procedure is as follows: one surrounds each lattice site by a ``cell'' of sizes $a^3$, where $a$ is the lattice spacing, and associates the charge of every test particle momentarily contained in the cell with this lattice site.  Currents are induced when test particles transit from one cell to the next; these lead to discontinuous charges of the electrical fields on the link orthogonal to the penetrated cell face. 

There are many technical details associated with the particle dynamics, which cannot all be discussed here.\cite{MHM98}  Suffice it to say that electric fields act on the test particles only when they switch between cells, while magnetic fields act at each time step.  If a particle does not have enough kinetic energy to transit into the next cell, it is reflected at the cell boundary.  Careful ordering of the updates of the various quantities at each time step insures parity and time-reversal invariance, and an appropriate leap-frog algorithm conserves the phase-space measure. 

\section{Results}

We have performed calculations on lattices as large as $32^3$ with several hundred test particles per cell.\cite{MHM98,Hu98}  Energy conservation can be achieved at the level of $10^{-5}$, and Gauss' law is conserved up to $10^{-10}$.  The parameters that can be freely chosen are: the gauge coupling $g$, the magnitude of the test particle color charge $Q$, the temperature $T$ of their momenum distribution, and the average number of test particles per cell $(na^3)$.  The HTL effective action contains these parameters only in the combination of the plasma frequency 
\begin{equation} 
  \omega_p^2  \sim g^2Q^2 n/T. 
\end{equation} 
The independent variability of these parameters in the test particle formulation allows, in practice, for a numerical test of the correct scaling behavior.  Let us now discuss some results. 

\begin{figure}[htb]
\vfill
\centerline{
\begin{minipage}[t]{.47\linewidth}\centering
\mbox{\epsfig{file=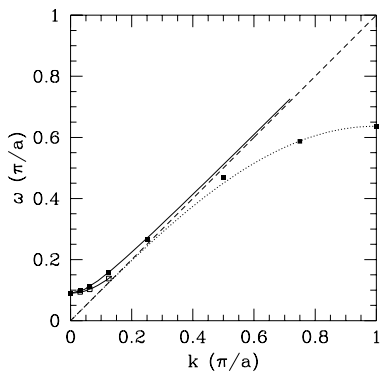,width=.99\linewidth}}
\end{minipage}
\hspace{.06\linewidth}
\begin{minipage}[b]{.47\linewidth}\centering
\caption{
Dispersion relation: free field in the continuum (dashed line: 
$\omega = k$); free field on the lattice (dotted line: 
$\omega a = 2\sin{(k a/2)}$); plasma modes: transverse (theory:
the upper solid curve; numerical results: solid rectangles) and longitudinal
(theory: the lower solid curve; numerical results: open rectangles).
[From ref.\protect\cite{MHM98}]}
\end{minipage}}
\end{figure}

The dispersion relation obtained on a $16^3$ lattice is shown in Fig.\ 1.  
At low wave numbers $k$, the eigenfrequencies nicely follow the expectation from continuum HTL perturbation theory (solid lines), whereas at large $k$, the frequencies approach the free lattice dispersion curve (dotted line).  For 
$k \neq 0$, longitudinal and transverse modes clearly have different frequencies, and the plasma frequency $w_p$ is obtained for $k = 0$.  The dependence of $w_p$ on the combination $gQ\sqrt{n/T}$ agrees very nicely with the HTL 
prediction, as shown in Figure 2.  The non-Abelian plasma modes are strongly damped.  Again, the analysis shows that the plasmon damping rate $\gamma_p$ for $k = 0$ is in excellent agreement with HTL perturbation theory (see Fig.\ 3). 

\begin{figure}[htb]
\vfill
\centerline{
\begin{minipage}[t]{.47\linewidth}\centering
\mbox{\epsfig{file=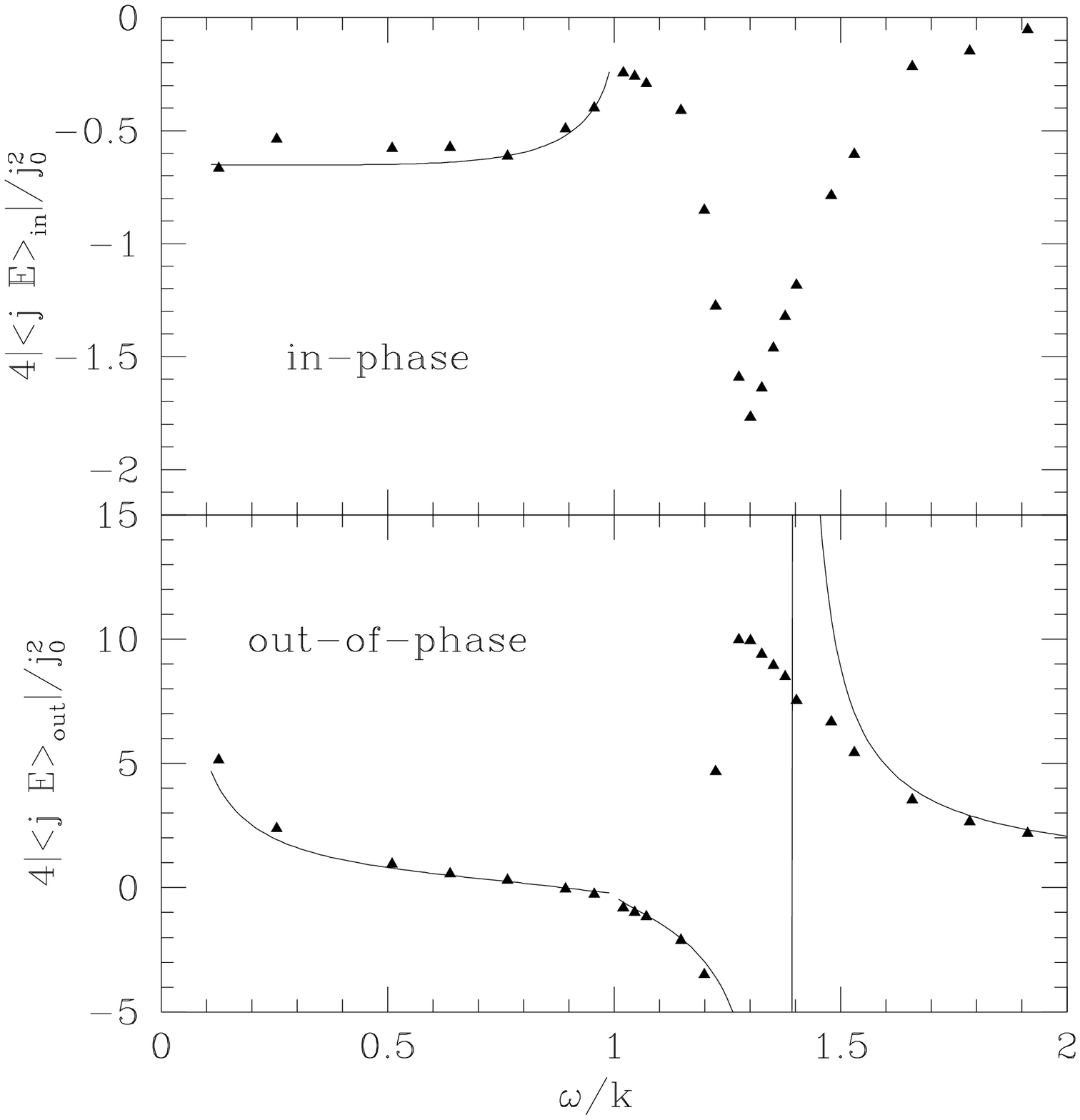,width=.99\linewidth}}
\caption{
Space-time average of $\vec{j}\cdot\vec{E}$ plotted as a function of
$\omega/k$ (longitudinal polarization).
Upper window: in-phase average;
lower window: out-of-phase average.
[From ref.\protect\cite{MHM98}]}
\end{minipage}
\hspace{.06\linewidth}
\begin{minipage}[t]{.47\linewidth}\centering
\mbox{\epsfig{file=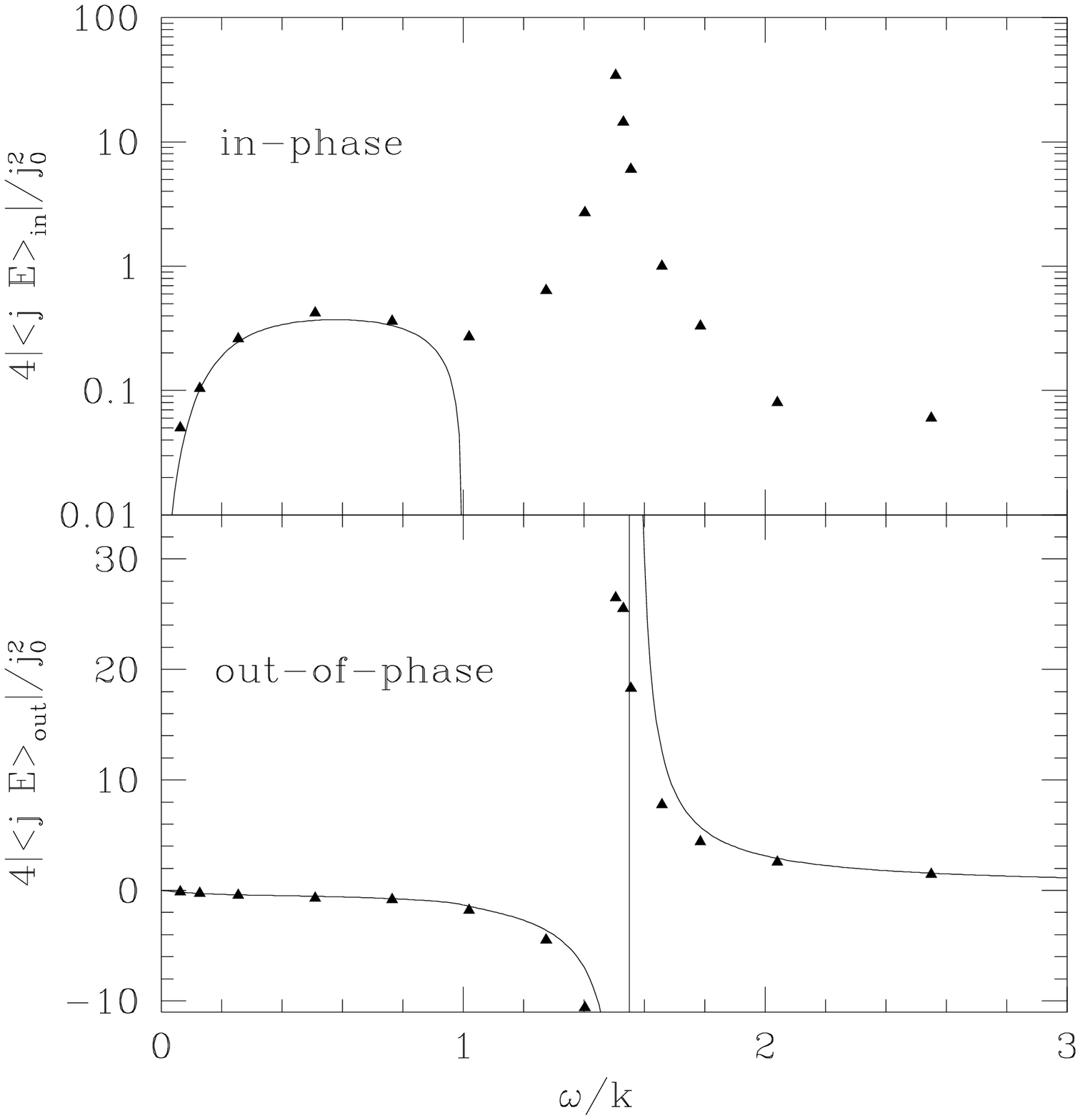,width=.99\linewidth}}
\caption{
Space-time average of $\vec{j}\cdot\vec{E}$ plotted as a function of
$\omega/k$ (transverse polarization).
Upper window: in-phase average;
lower window: out-of-phase average.
[From ref.\protect\cite{MHM98}]}
\end{minipage}}
\end{figure}

While this is encouraging, it is important for many applications to verify the correct HTL behavior of the response functon for arbitrary combinations of $\omega$ and $k$.  In the Abelian limit, this can be achieved by a linear response analysis, when the system of gauge field and test particles is driven by a weak current of the form 
\begin{equation} 
\vec{j} (\vec{x}, t) 
= \hat{j} j_0 \sin (\omega t) \sin \left(\vec{k} \cdot \vec{x}\right). 
\end{equation} 
The dissipative and dispersive part of the response function can then be obtained by means of the in-phase and out-of-phase averages: 
\begin{equation} 
\langle \vec{j} (x, t) \cdot \vec{E} (x, t) \rangle \quad \mbox{ and } \quad \langle \vec{j} (x, t) \cdot \vec{E} (x, t - \frac{\pi}{2\omega}) \rangle, 
\end{equation} 
respectively.  Transverse and longitudinal response are simply obtained by choosing $\hat j \perp \vec{k}$ or $\hat{j} \parallel \vec{k}$.  Again, one finds nice agreement with the HTL predictions for both response functions, 
as shown in Figs.\ 4, 5.  The spikes of the response functions
 at the plasma resonance frequency $\omega_p (k)$ are clearly visible.  The smallness of the dispersive  part of the transverse response for $\omega < k$ is indicative of the absence of magnetic screening at the HTL level. 

\begin{figure}[htb]
\vfill
\centerline{
\begin{minipage}[t]{.47\linewidth}\centering
\mbox{\epsfig{file=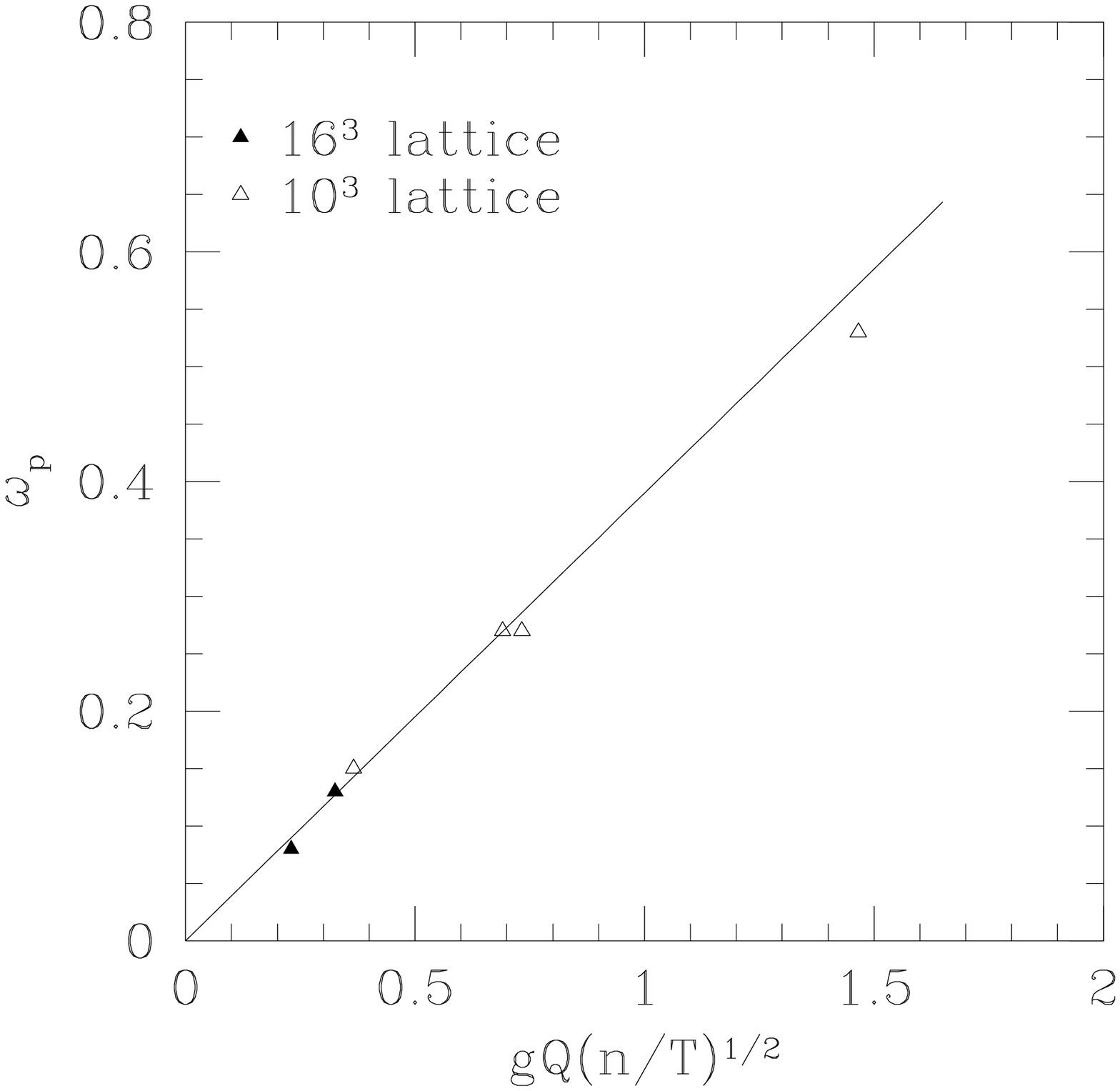,width=.99\linewidth}}
\caption{
Scaling behavior of the plasma frequency.
Solid line: $\omega_p = 0.39 gQ \sqrt{{\langle n \rangle}/T}$ (theory);
triangles: numerically obtained values.
[From ref.\protect\cite{Hu98}]}
\end{minipage}
\hspace{.06\linewidth}
\begin{minipage}[t]{.47\linewidth}\centering
\mbox{\epsfig{file=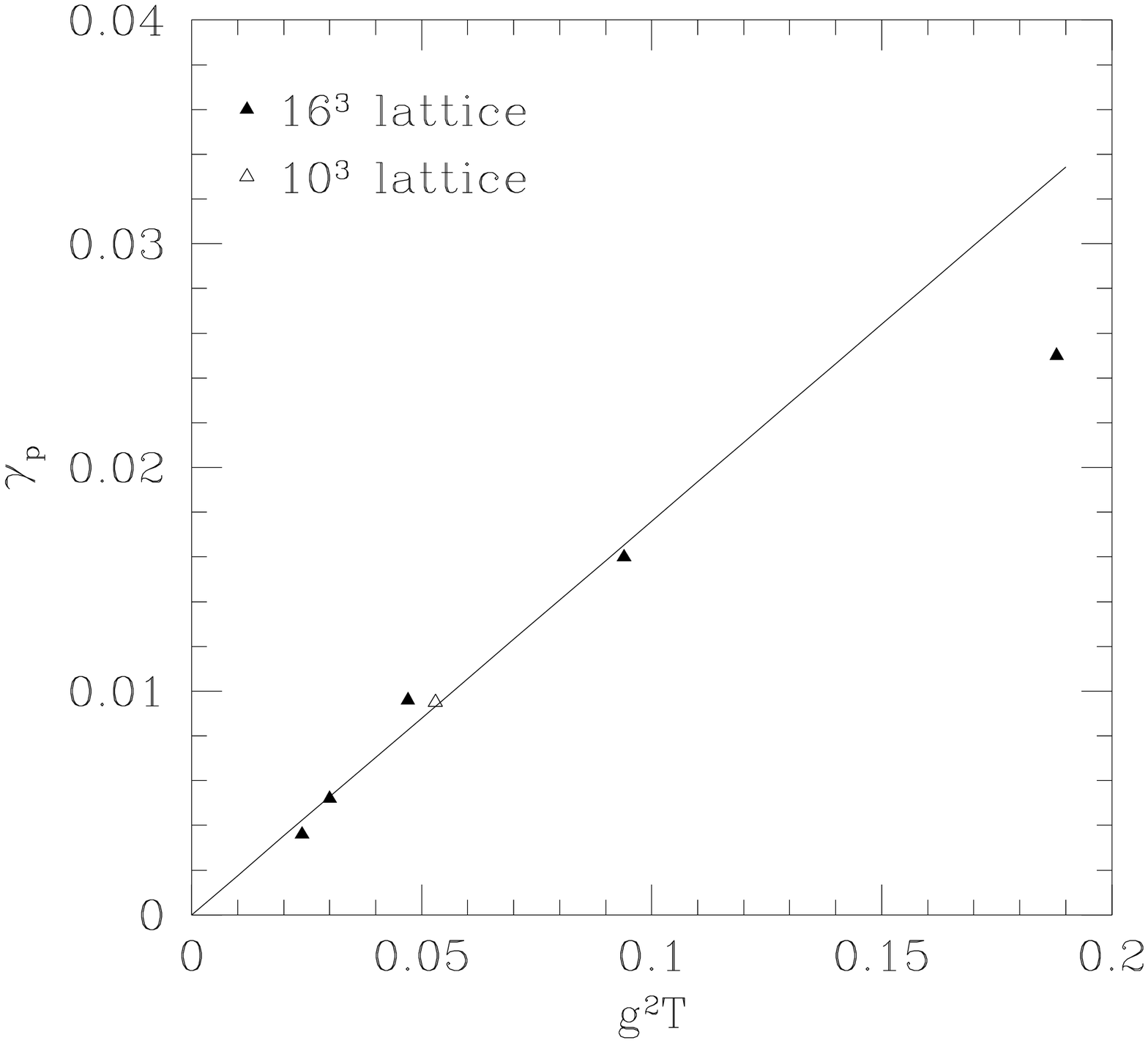,width=.99\linewidth}}
\caption{
Scaling behavior of the plasma rates.
Solid line: $\gamma_p = 0.176 g^2 T$ (theory);
triangles: numerically obtained values.
[From ref.\protect\cite{Hu98}]}
\end{minipage}}
\end{figure}

As an application, we have studied the diffusion rate of the Chern-Simons number $N_{\rm CS}$ in the 
unbroken $SU(2)$-Higgs Theory, which corresponds to the high-temperature phase of the electroweak theory in the limit $\sin ^2\theta_W \to 0$.  It was long suspected, on the basis of scaling arguments that this rate $\Gamma_{\rm CS}$ should be proportional to $(g^2T)^4$.   However, Arnold, Son, and Yaffe argued convincingly that the rate should be suppressed by another factor $g^2$, because space-like modes of the gauge field are strongly damped by the HTL dynamics.  Since 
\begin{equation}   
\Pi_{\rm HTL} (k, \omega) \stackrel{\omega \ll k}{\longrightarrow} - i (gT)^2 \frac{\omega}{k} 
= -3i \omega^2_p \frac{\omega}{k}
\end{equation}  
the relevant frequency domain for long-distance fluctuations of the gauge field becomes 
\begin{equation} 
\omega \sim  \frac{k^3}{\omega^2_p} ~ g^4T \quad (\mbox{ for } k \sim g^2T). 
\end{equation} 
Our numerical simulations of $\Gamma_{\rm CS}$ show that, indeed, the rate depends linearly on $1/\omega^2_p$ as expected from (17), 
confirming the ASY scaling law (see Fig. 6).\footnote{More recently, 
B\"odecker has argued that the dissipative response of the hard thermal loops has logarithmic corrections  $\sim\log (1/g)$.  We observed no indication for such a behavior, but the logarithm may be difficult to see in the numerical results.}

\begin{figure}[htb]
\vfill
\centerline{
\begin{minipage}[t]{.47\linewidth}\centering
\mbox{\epsfig{file=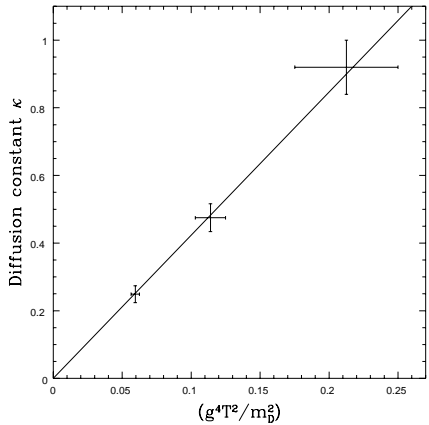,width=.99\linewidth}}
\end{minipage}
\hspace{.06\linewidth}
\begin{minipage}[b]{.47\linewidth}\centering
\caption{
Result for $\kappa$ at three values of average particle number 
$\langle n \rangle$ but a common value of lattice spacing and other 
physical parameters ($g$, $Q$, $T$), plotted against $g^4 T^2/m_{\rm D}^2$.  
The error bars in $m_{\rm D}^2$ reflect uncertainty in the damping from 
hard classical lattice modes.  The old picture predicts a flat line, 
while the ASY picture predicts a straight line through the origin, like the 
illustrated fit.
[From ref.\protect\cite{MHM98}]}
\end{minipage}}
\end{figure}

\section{Summary}

The infrared limit of thermal gauge field dynamics in Minkowski space can be represented by the local interaction of the classical gauge field with a thermal plasma of classical colored particles. A lattice formulation of this dynamical system can be derived which conserves all important symmetries and lends itself to numerical simulation. We have verified that the HTL dispersion relations known from perturbation theory are well reproduced, and we have confirmed the Arnold-Son-Yaffe scaling relation for the Chern-Simons number diffusion rate in the Yang-Mills-Higgs model. 

Two interesting generalizations for future work are: The study of the evolution of gauge field configurations far off equilibrium, such as they occur in relativistic heavy ion reactions, and the inclusion of hard interactions among the colored particles, generalizing the non-Abelian Vlasov equation to a non-Abelian Boltzmann equation and effectively introducing a color mean field into the parton cascade model.\cite{Poschl}

\section*{Acknowledgments}
This work was supported in part by grant no. DE-FG02-96ER40945 from the U.S.\ Department of Energy.


\section*{References}

\begin{thebibliography}{99}

\bibitem{BP90} E. Braaten and R.D. Pisarski, 
        {\em Nucl. Phys.} {\bf B 337} (1990) 569;
	{\em Phys. Rev.} {\bf D 42} (1990) 2156.

\bibitem{TW90} J.C. Taylor and S.M.H. Wong, 
        {\em Nucl. Phys.} {\bf B 346} (1990) 115.

\bibitem{BI93} J.P. Blaizot and E. Iancu, 
        {\em Phys. Rev. Lett.} {\bf 70} (1993) 3376.

\bibitem{Wong70}
        S.K. Wong, {\em Nuovo Cim.} {\bf A 65} (1970) 689.

\bibitem{Heinz83} U. Heinz, 
       {\em Phys. Rev. Lett.} {\bf 51} (1983) 351;
       {\em Ann. Phys. (NY)} {\bf 161} (1985) 48;
       {\bf 168} (1986) 148.

\bibitem{KLLM94} P.F. Kelly, Q. Liu, C. Lucchesi, and C. Manuel, 
        {\em Phys. Rev. Lett.} {\bf 72} (1994) 3461;
        {\em Phys. Rev.} {\bf D 50} (1994) 4209.

\bibitem{HM96} C.R. Hu and B. M\"uller, 
        {\em Phys. Lett.} {\bf B409} (1997) 377.

\bibitem{MHM98} G.D. Moore, C.R. Hu and B. M\"uller, 
        {\em Phys. Rev. D} {\bf 58} (1998) 045001.

\bibitem{Hu98} C.R. Hu, Ph.D. thesis (unpublished), Duke University, 1998.

\bibitem{Poschl} S.A. Bass, B. M\"uller, and W. P\"oschl,
        preprint nucl-th/9808011; 
        B. M\"uller, and W. P\"oschl, preprint nucl-th/9808031.

\end{thebibliography}
\end{document}